\documentclass[epj]{svjour}
\usepackage{graphics}
\usepackage{amsmath}
\usepackage{amssymb}

\begin{document}

\title{Analysis of the quadrupole deformation of $\Delta$(1232)
within an  effective Lagrangian model
for pion photoproduction from the nucleon}
\author{C. Fern\'andez-Ram\'{\i}rez\inst{1,2} 
\thanks{Corresponding author; \email{cesar@nuc2.fis.ucm.es}}
\and E. Moya de Guerra\inst{1,3}
\and J.M. Ud\'{\i}as\inst{3}} 

\institute{
Instituto de Estructura de la Materia,
CSIC, Serrano 123, E-28006 Madrid, Spain
\and
Departamento de F\'{\i}sica At\'omica, Molecular y Nuclear,
Universidad de Sevilla, Apdo. 1065, E-41080 Sevilla, Spain
\and
Departamento de F\'{\i}sica 
At\'omica, Molecular y Nuclear, Facultad de 
Ciencias F\'{\i}sicas, Universidad Complutense de Madrid, 
Avda. Complutense s/n, E-28040 Madrid, Spain}

\date{Received: date / Revised version: date}

\abstract{
We present an extraction of the E2/M1 ratio of the $\Delta$(1232)
from experimental data applying an effective Lagrangian model.
We compare the result obtained with different nucleonic models
and we reconcile the experimental results with the Lattice QCD
calculations.
}

\titlerunning{Analysis of the quadrupole deformation of $\Delta$(1232)\dots}
\maketitle

The deformation of the nucleon and its first
excitation, the $\Delta(1232)$, 
is a topic that has focused the attention of many
researchers in the last years from both the experimental and the theoretical
sides \cite{KS}. 
The possibility of such deformation has been studied using the E2/M1
ratio (EMR) of the $\gamma N \to \Delta(1232)$ transition \cite{bernstein}.
The emission (absorption) of a photon by a spin-3/2 particle involves a 
magnetic dipole (M1) multipolarity and an electric quadrupole (E2) 
multipolarity. 
From experiments it is found that E2 is small but not zero, which evokes
a deformed nucleon picture.
A deviation from zero of the EMR is a clear indication of 
the existence of such deformation and allows to quantify it. This
ratio is mainly obtained in two different ways, from nucleonic models such as
quark models or Lattice QCD, and from experimental data.  
Reconciliation of both
extractions is significant in order to understand the structure of the nucleon.
This work is concerned with the extraction of the intrinsic 
E2/M1 ratio of the $\Delta$(1232) from experimental
data using a reaction model. 

In \cite{phd,fernandez06a} 
we have developed a pion photoproduction model up to 
1 GeV of photon energy based upon effective Lagrangians.
The model follows closely the work of Garcilazo and Moya de Guerra
\cite{EMoya} and has similarities with the work of Sato and
Lee \cite{Sato01} as well as with other works 
\cite{Davidson91,Pasc04,Vanderhaeghen} 
based on the seminal work of Peccei \cite{Peccei}.
The reaction model allows us to isolate the contribution of the
$\Delta$(1232) -- taking into account the high energy behaviour 
of the tail of the resonance --, 
to calculate its EMR, and to compare it with the values provided by
nucleonic models. 
In addition to Born terms (those which involve only photons, 
nucleons, and pions) and vector meson exchange terms 
($\rho$ and $\omega$ exchanges), the model includes 
all the four star resonances in Particle Data Group (PDG)
\cite{PDG2006} up to 1.7 GeV mass
and up to spin-3/2:
$\Delta$(1232), N(1440), N(1520), $\Delta$(1620), N(1650), 
and $\Delta$(1700).

The model displays chiral symmetry, gauge invariance,
and crossing symmetry, as well as a consistent treatment
of the interaction with spin-3/2 particles that avoids
well-known pathologies present in previous models 
\cite{phd,fernandez06a}.
The dressing of the resonances is considered by means of a 
phenomenological width which takes into account 
decays into one $\pi$, one $\eta$, and two $\pi$.
The width fulfills crossing symmetry and contributes
to both direct and crossed channels of the resonances.

We assume that the final state interactions (FSI) factorize 
($\pi N$ rescattering)
and can be included through the distortion of the 
$\pi N$ final state wave function.
The calculation of the distortion requires one to  
calculate higher order pion loops or to develop a 
phenomenological potential FSI model.
Both approaches are far of the one we apply.
The first one is overwhelmingly complex
and the second would introduce additional model-dependencies,
which are to be avoided in the present analysis, 
because we are mainly interested in the bare properties 
of the resonances.
We rather include FSI in a phenomenological way by
adding a phase $\delta_{\text{FSI}}$ to the electromagnetic multipoles.
We determine this phase so that the total phase of the electromagnetic
multipole is identical to the one of the  
energy dependent solution of SAID \cite{SAID}.
In this way we are able to isolate the 
electromagnetic vertex and remove the FSI effects.

\begin{figure*}
\begin{center}
\rotatebox{-90}{\scalebox{0.40}[0.45]{\includegraphics{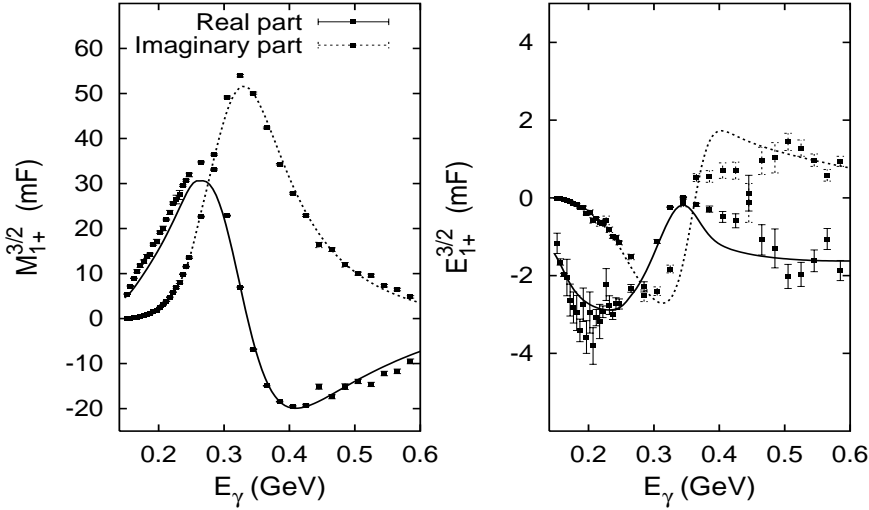}}}
\caption{$M_{1+}^{3/2}$ and $E_{1+}^{3/2}$ electromagnetic multipoles.
Curve conventions: solid, real part of the multipole; dashed, 
imaginary part of the multipole. Data taken from \cite{SAID}.}
\label{fig:1}
\end{center}
\end{figure*}

In order to assess the parameters of the model we had to
minimize the function $\chi^2$ defined by
\begin{equation}
\chi^2 = \sum_{j=1}^m \left[\frac{\mathcal{M}^{exp}_j-
\mathcal{M}^{th}_j \left( \lambda_1, \dots, \lambda_n \right)}
{\Delta \mathcal{M}^{exp}_j} \right]^2,
\end{equation}
where ${\mathcal M}^{exp}$ stands for the current energy 
independent extraction of the multipole analysis of SAID up 
to 1 GeV for $E_{0+}$, $M_ {1-}$, $E_{1+}$, $M_{1+}$, $E_{2-}$, 
and $M_{2-}$ multipoles in the three isospin channels 
$I=\frac{3}{2},p,n$ for the $\gamma p \to \pi^0 p$ 
process \cite{SAID}.
$\Delta \mathcal{M}^{exp}$ is the error 
and $\mathcal{M}^{th}$ is the multipole given by the model 
which depends on the parameters $\lambda_1$, $\dots$, $\lambda_n$,
which stand for the electromagnetic coupling constans of the resonances
and the cutoff $\Lambda$ which regularizes the high energy behaviour 
of the Born terms. The masses and the widths of the resonances
have been taken from the multichannel analysis of Vrana, Dytman, and Lee
\cite{Vrana} and from a \textit{speed plot} calculation \cite{fernandez06a}.
The EMR we present has been obtained as an average of the results obtained
with both sets of masses and widths.
Electromagnetic multipoles are complex quantities and
we have taken into account 763 data for the real part of 
the multipoles and the same amount for the imaginary part. 
Thus, $m=1526$ data points have been used in the fits.

In order to fit the data and determine 
the best parameters of the resonances 
we have written a genetic algorithm combined with the 
\texttt{E04FCF} routine from NAG libraries \cite{NAG}. 
Although genetic algorithms are computationally more expensive 
than other algorithms, in a minimisation problem it is much 
less likely for them to get stuck at local minima than for other
methods, namely gradient based minimisation methods. 
Thus, in a multiparameter minimisation like the one we face here
it is probably the best possibility to search for the minimum
\cite{phd,Ireland}. In Fig. \ref{fig:1} we show the fits to
$M_{1+}^{3/2}$ and $E_{1+}^{3/2}$ electromagnetic multipoles.

Different definitions of the EMR have been employed
in the literature. We should distinguish between 
the \textit{intrinsic} or \textit{bare} EMR of the $\Delta$(1232) 
and the directly measured 
value in experiments
which is often called \textit{physical} or \textit{dressed} 
EMR value \cite{fernandez06b,Pasc04,Sato01}.
The physical EMR is obtained as the ratio between the 
imaginary parts of $E_{1+}^{3/2}$
and $M_{1+}^{3/2}$ electromagnetic multipoles
at the invariant mass (photon energy) at which 
$\text{Re}\left[ M_{1+}^{3/2}\right]=0=\text{Re}\left[E_{1+}^{3/2}\right]$.
Since all the reaction models are fitted to the experimental electromagnetic
multipoles, they generally reproduce the physical EMR value
within the error bars (see table \ref{tab:experiments}). We obtain 
\begin{equation}
\text{EMR}^{\text{physical}}
=\frac{\text{Im}\left[E_{1+}^{3/2}\right]}{\text{Im} 
\left[M_{1+}^{3/2}\right]} \times 100 \%
=\left( -3.9 \pm 1.1 \right) \%.
 \end{equation}

However, this measured EMR value is not directly available 
from theoretical models of the nucleon and its resonances. 
Instead, if we want to compare to models of nucleonic structure, 
it is necessary to extract the bare EMR value
of $\Delta$(1232) which is defined as
\begin{equation}
\text{EMR}^{\text{bare}}
=\frac{G_E^{\Delta(1232)}}{G_M^{\Delta(1232)}}\times 100\%
=\left( -1.30 \pm 0.52 \right) \%, \label{eq:EMR}
\end{equation}

The EMR defined in this way
depends only on the intrinsic characteristics 
of the $\Delta$(1232) and can thus
be compared directly to predictions from  nucleonic models. 
It is not, however,
directly measurable but must be inferred (in a model dependent way) 
from reaction models.

The intrinsic quadrupole deformation of the $\Delta$(1232) is found to be
EMR$=\left( -1.30\pm0.52 \right) \%$, indicative of a small
oblate deformation. 
In Tables \ref{tab:experiments} and \ref{tab:models} 
we compare our EMR values (bare and physical) to
the ones extracted by other authors using other models for pion
photoproduction, as well as to 
predictions of nucleonic models.

\begin{table}
\caption{Comparison of EMR$^{\text{physical}}$ 
values from experiments compared to the 
values obtained with reaction models.} \label{tab:experiments}
\begin{tabular}{lll}
\hline\noalign{\smallskip}
          & EMR$^{\text{physical}}$ & Ref.\\ 
\noalign{\smallskip}\hline\noalign{\smallskip}
\textbf{Experiments} & & \\
LEGS Collaboration & 
$\left(-3.07 \pm 0.26 \pm 0.24 \right) \%$ 
& \cite{Blanpied} \\
A1 Collaboration & 
$ \left(-2.28\pm 0.29 \pm 0.20 \right) \%$ & \cite{Stave} \\
A2 Collaboration &
$ \left(-2.74\pm 0.03 \pm 0.30 \right) \%$ & \cite{Ahrens} \\
Particle Data Group 
& $\left(-2.5 \pm 0.5 \right) \%$ & \cite{PDG2006} \\
\textbf{Reaction models}& & \\ 
Fern\'andez-Ram\'{\i}rez \textit{et al.}&
$\left( -3.9 \pm 1.1 \right) \%$& \cite{fernandez06b}\\
Pascalutsa and Tjon  &$\left( -2.4\pm0.1 \right)$\%   & \cite{Pasc04} \\
Sato and Lee         &$-2.7$\%        & \cite{Sato01} \\
Fuda and Alharbi     &$-2.09$\%       & \cite{Fuda} \\
\noalign{\smallskip}\hline
\end{tabular}
\end{table}

Reconciliation of the experimental value of the 
E2/M1 $\gamma N \to \Delta(1232)$ transition ratio 
(EMR$^{\text{physical}}$)
with the one 
obtained using Lattice QCD (EMR$^{\text{bare}}$ ) 
within a consistent and sound framework
(besides our analysis only in \cite{Pasc04} 
a consistent
treatment of the spin-3/2 interaction is performed).
is one of the goals of this work \cite{fernandez06b}. 
Our results also indicate that quark models
need improvements in order to reproduce the value obtained from experiment.

\begin{table}
\caption{Comparison of EMR$^{\text{bare}}$ 
values extracted from experiments through 
reaction models compared to the 
values obtained with nucleonic models.} \label{tab:models}
\begin{tabular}{lll}
\hline\noalign{\smallskip}
          & EMR$^{\text{bare}}$ & Ref.\\ 
\noalign{\smallskip}\hline\noalign{\smallskip}
\textbf{Reaction models} & & \\
Fern\'andez-Ram\'{\i}rez \textit{et al.}&
$\left( -1.30\pm0.52 \right) \%$
& \cite{fernandez06b}\\
Pascalutsa and Tjon  &$\left( 3.8\pm1.6 \right)$\%   & \cite{Pasc04} \\
Sato and Lee         &$-1.3$\%        & \cite{Sato01} \\
Davidson \textit{et al.} & $-1.45$\% & \cite{Davidson91} \\
Garcilazo and Moya de Guerra &$-1.42$\% & \cite{EMoya} \\
Vanderhaeghen \textit{et al.}
&$-1.43$\% & \cite{Vanderhaeghen} \\
\textbf{Nucleonic models}& & \\ 
Non-relativistic quark model& 0\% & \cite{Becchi}\\
Constituent quark model &$-3.5$\% & \cite{Buchmann} \\
Skyrme model& $\left( -3.5\pm1.5 \right) $\%& \cite{Wirzba}\\
Lattice QCD (Leinweber \textit{et al.})& $\left( 3 \pm 8 \right)$\%
& \cite{Leinweber}\\
Lattice QCD (Alexandrou \textit{et al.})& &\cite{Alexandrou} \\
($Q^2=0.1$ GeV$^2$, $m_\pi = 0$)
&$\left( -1.93 \pm 0.94 \right)$\%& \\
($Q^2=0.1$ GeV$^2$, $m_\pi = 370$ MeV)
&$\left( -1.40 \pm 0.60 \right)$\%& \\
\noalign{\smallskip}\hline
\end{tabular}
\end{table}

\begin{acknowledgement}
C.F.-R. work has been developed under Spanish 
Government grant UAC2002-0009. This work
has been supported in part under contracts of 
Ministerio de Educaci\'on y Ciencia (Spain) 
FIS2005-00640 and BFM2003-04147-C02-01.
\end{acknowledgement}

\end{document}